\documentclass[aps,pre,preprint,onecolumn,floatfix,12pt]{revtex4-1}
\usepackage{subfigure,amssymb,amsmath,natbib,mathtools} 
\usepackage[utf8]{inputenc}

\usepackage{textcomp} 

\usepackage{graphicx}
\usepackage{dcolumn}
\usepackage{bm}
\usepackage[usenames]{color}
\usepackage [babel=true]{csquotes}
\usepackage{MnSymbol}
\definecolor{mygray}{gray}{0.5}
\usepackage{enumerate}
\usepackage{chemfig}
\usepackage{wrapfig}
\usepackage{color}

\begin{document}

\title{Surface tension and the origin of the circular hydraulic jump in a thin liquid film}

\author{Alexis Duchesne, Anders Andersen and Tomas Bohr}

\affiliation{Department of Physics, Technical University of Denmark, DK-2800 Kgs. Lyngby, Denmark}

\date{\today}

\begin{abstract}
It was recently claimed by Bhagat {\em et al.} (J. Fluid Mech. vol. 851 (2018), R5) that the scientific literature on the circular hydraulic jump in a thin liquid film is flawed by improper treatment and severe underestimation of the influence of surface tension. Bhagat {\em et al.} use an energy equation with a new surface energy term that is introduced without reference, and they conclude that the location of the hydraulic jump is determined by surface tension alone. We show that this approach is incorrect and derive a corrected energy equation. Proper treatment of surface tension in thin film flows is of general interest beyond hydraulic jumps, and we show that the effect of surface tension is fully contained in the Laplace pressure due to the curvature of the surface. Following the same approach as Bhagat {\em et al.}, i.e., keeping only the first derivative of the surface velocity, the influence of surface tension is, for thin films, much smaller than claimed by them. We further describe the influence of viscosity in thin film flows, and we conclude by discussing the distinction between time-dependent and stationary hydraulic jumps.
\end{abstract}

\maketitle
\section{Introduction}

The circular hydraulic jump appears when a vertical liquid jet impinges on a horizontal plate and results in a rotationally symmetric outward flow. It is well known that such a flow, if it is rapid enough, gives rise to a circular hydraulic jump, where the height of the liquid layer increases abruptly, with a corresponding decrease in radial velocity. The origin of such jumps is typically explained as a transition from supercritical to subcritical flow, such that the relations between the heights and velocities are governed by the Rayleigh-B\'elanger condition and thus influenced by gravity \cite{Watson1964, Bohr1993, Bush2003, Duchesne2014}, whereas surface tension plays a less significant role \cite{Bush2003}. It is also known that flow separation is closely linked to the jump, so that the jump can develop due to an adverse pressure gradient, possibly caused by other forces than gravity \cite{Tani1949, Craik1981, Bohr1996, Watanabe2003}.

\begin{figure}[!h]
\begin{centering}
\includegraphics[width=0.85\textwidth]{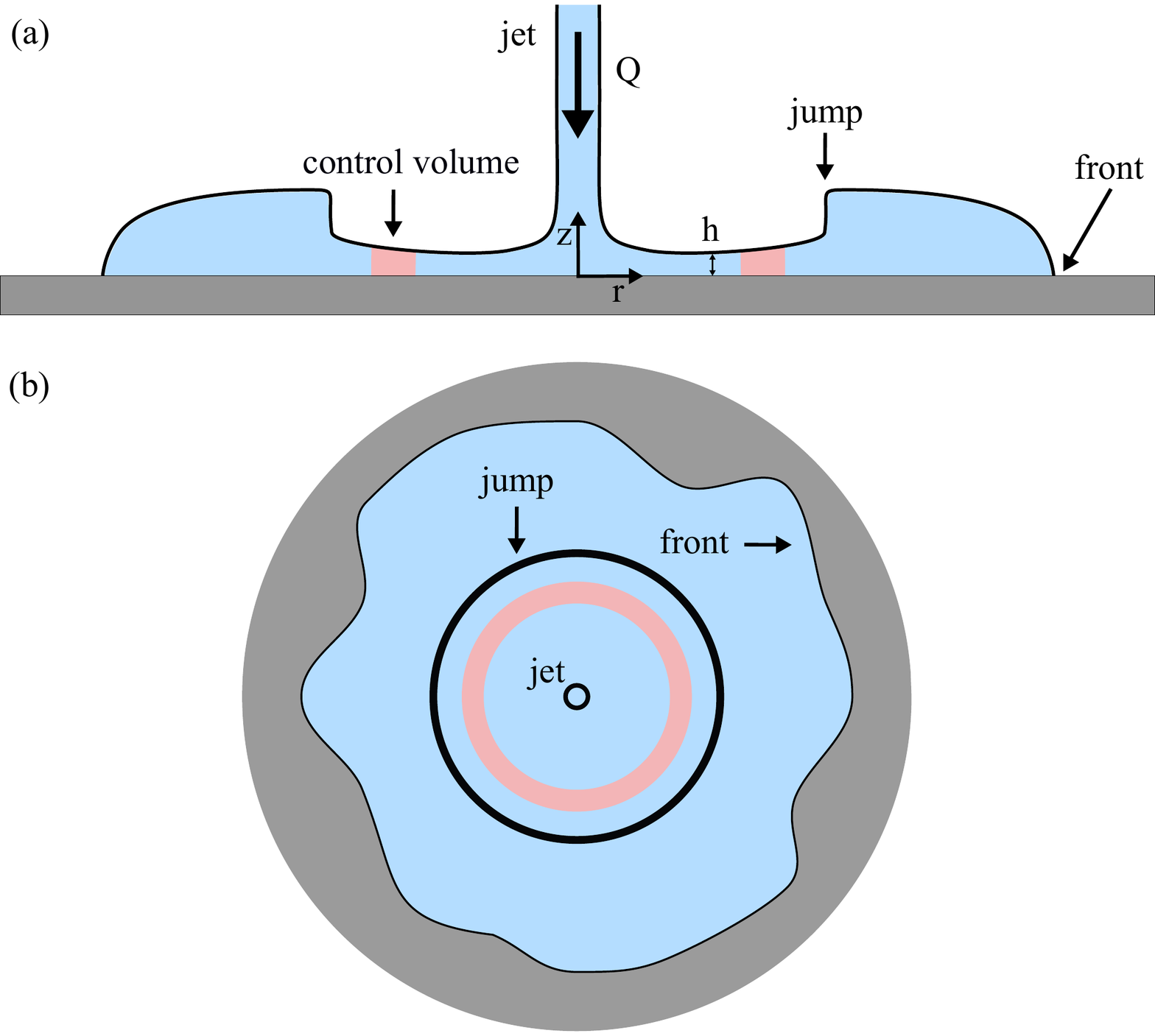}
\caption{\label{fig-geometry} Schematic illustration of the circular hydraulic jump appearing when a liquid (blue) impinges in a jet on a horizontal plate (grey). (a) Side-view of a cut through the liquid layer in a vertical plane containing the axis of symmetry. The positions of the jump and the expanding liquid front (contact line) are indicated. (b) Top-view that shows the finite extent of the circular bottom plate and the irregular shape of the expanding liquid front. Both views show the fixed annular control volume (red) inside the jump region.}
\end{centering}
\end{figure}

In a recent paper, \citet{Bhagat2018} investigated the origin of the circular hydraulic jump in a thin liquid film with a free surface and claimed that surface tension plays a much larger role than hitherto believed. In contrast to the above mentioned papers, the authors state that they are not studying stationary hydraulic jumps. Instead, they are looking at the jump while the liquid front is expanding and has not yet reached the edge of the plate. As shown in their videos, there is an intermediate time interval in which the jump position is practically constant, although the exterior (non-circular) front is still expanding as shown schematically in Fig.~\ref{fig-geometry}. How this is reflected in their theory is not clear, since all of their equations are {\em time-independent}. Similarly, the annular control volume used to derive their energy equation is at a fixed position inside the jump region (Fig.~\ref{fig-geometry}), and the surface energy that the authors introduce is not related to the expansion of the liquid surface outside the jump.

Following \citet{Bhagat2018} we consider a time-independent, rotationally symmetric flow created by a central jet impinging normally on a flat solid surface. We characterize the flow by the locally defined dimensionless film thickness (aspect ratio), Weber number, and Froude number

\begin{equation}
\label{eq:dimensionless}
   \alpha      \ = \ \frac{h}{r} \ , \hspace{1cm} 
   \mathrm{We} \ = \ \frac{\rho \, u_{\mathrm{s}}^2 \, h}{\gamma} \ , \hspace{1cm}     
   \mathrm{Fr} \ = \ \frac{u_{\mathrm{s}}}{\sqrt{g \, h}} \ , 
\end{equation}

\noindent
where $r$ denotes the radial coordinate, $h$ the thickness of the liquid film, $u_{\mathrm{s}}$ the radial flow velocity at the free surface, $\rho$ the density, $\gamma$ the surface tension, and $g$ the acceleration due to gravity. \citet[Eq. (5.8)]{Bhagat2018} claim that the radius of the hydraulic jump is determined by a condition of the form

\begin{equation}
\label{R1}
{\rm We}^{-1} + {\rm Fr}^{-2} \ \approx \ 1 \ ,
\end{equation}
whereas we shall argue that a correct derivation leads to  the condition
\begin{equation}
\label{R2}
\alpha^2 \, {\rm We}^{-1} + {\rm Fr}^{-2} \ \approx \ 1 \ .
\end{equation}
In thin films the aspect ratio, $\alpha$, is much smaller than unity and the two conditions are qualitatively different.  The condition (\ref{R2}) is in agreement with the theoretical model by \citet{Mathur2007}, that was derived by averaging the Navier-Stokes equation through the boundary layer in the inner flow region. Instead of working directly with the Navier-Stokes equation, \citet{Bhagat2018} derived the condition (\ref{R1}) by introducing an energy equation that includes a ``new term" representing ``the flux of surface energy that has been neglected in previous studies". Here, we show that this energy equation is in error, and that it is in disagreement with fundamental fluid dynamical theory.  We should note from the outset that neither of the two conditions can be taken generally as a prediction of the location of the hydraulic jump. These conditions are found from estimates of where the flow inside the hydraulic jump would become singular, obtained by lowest order approximation, retaining only the first order derivatives of $u_{\mathrm{s}}$ or $h$, as will be explained in the following. The position of the hydraulic jump will, in general, depend both on the inner and outer flow, so conditions like (\ref{R1}) or (\ref{R2}) can at best represent upper bounds on the jump-radius.

\section{Interfacial condition on normal and tangential stress}

The issue is how to include surface tension in the description of a flowing liquid with a free surface. Surface tension does not enter the Navier-Stokes equation directly, and it appears only in the interfacial condition on the stress at the free surface. In a Newtonian and incompressible flow with pressure $p$ and velocity field $v_{i}$, the stress tensor is the sum of the pressure term and the viscous term

\begin{equation}
\sigma_{ij} \ = \ - p \, \delta_{ij} \, + \, \mu \, v_{ij} \ ,
\end{equation}

\noindent
where $\mu$ is the viscosity and $v_{ij} \, = \, \partial v_{i}/\partial x_{j} \, + \, \partial v_{j}/\partial x_{i}$. At each point on the free surface the stress condition can be written as a tangential component

\begin{equation}
\label{eq:tangential}
v_{ij} \, t_i \, n_j \ = \ 0 \ ,
\end{equation}
 
\noindent
and a normal component
\begin{equation}
\label{eq:normal}
p \, - \, p_{\mu} \ = \ p_0 \, + \, p_{\gamma} \ ,
\end{equation}

\noindent
where $n_{i}$ and $t_{i}$ denote the normal and the tangential unit vector, respectively. The external pressure is $p_0$, the viscous pressure:
\begin{equation}
\label{pmu}
p_{\mu} \ = \ \mu \, v_{ij} \, n_i \, n_j \ =  \ 2 \, \mu \,  \frac{\partial v_{i}}{\partial x_{j}} \, n_i \, n_j \ ,
\end{equation}
and the Laplace pressure due to surface tension:

\begin{equation}
\label{pgamma}
p_{\gamma} \ = \ \gamma \left(\frac{1}{R_1}+ \frac{1}{R_2}\right) \ = \ \gamma \, \kappa \ ,
\end{equation}
where the curvature $\kappa$  is the sum of the principal normal curvatures $1/R_{1}$ and $1/R_{2}$. Surface tension thus only enters the normal component of the stress condition, and its effect is fully contained in the Laplace pressure. These results hold in general when the surface tension is uniform, both when the liquid is stationary and when it is flowing (see e.g., \citet{Batchelor1967}, pages 69 and 149-150). 

\section{Energy equation with surface tension}

Both \citet{Mathur2007} and \citet{Bhagat2018} consider the classical configuration in which a circular liquid jet impinges vertically down on a horizontal and flat solid surface and results in a rotationally symmetric thin film liquid flowing radially outwards, potentially leading to a circular hydraulic jump (Fig.~\ref{fig-geometry}). As shown there, the jump has formed and remains practically stationary while the outer front (contact line) is still moving out, typically in an irregular way dependent on the wetting properties of the solid surface (grey).
Both sets of authors investigate the inner flow using averaging theory and a self-similar horizontal velocity profile
\begin{equation}
\label{eq:profile}
u(r,z) \ = \ u_{\mathrm{s}}(r) f(\eta) \ , 
\end{equation}

\noindent
where $z$ is the vertical coordinate and $\eta = z/h(r)$. The corresponding vertical velocity field has the form
\begin{equation}
\label{eq:profilev}
w(r,z) \, = \, u_s(r) \, h'(r) \, \eta \, f(\eta), 
\end{equation}
as can be seen by using the incompressibility condition together with the kinematic boundary condition: $w = h' u$ at the free surface $z=h$.
Indeed such a self-similar inner flow solution exists when both gravity and surface tension are neglected \cite{Watson1964}. In the following we shall also assume the self-similar velocity profile (\ref{eq:profile})-(\ref{eq:profilev}), and in our equations we shall use the profile-dependent numerical coefficients 

\begin{equation}
C \ = \ \int_{0}^{1} (1 - \eta) \, f(\eta) \, \mathrm{d}\eta \ , \hspace{1cm} C_{n} \ = \ \int_{0}^{1} \left[f(\eta)\right]^{n} \, \mathrm{d}\eta \ . 
\end{equation}

\noindent
The energy balance, that is postulated in \citet[Eq. (5.2)]{Bhagat2018}, is written down for a fixed annular control volume with inner radius $r$ and outer radius $r + \Delta r$. The kinetic energy and surface tension terms of the equation are 
\begin{equation}
\label{E1}
        \frac12 \left. \left( \rho \, \bar{u^2} \, \bar{u} \, r \, h \right) \right|_r  
\, - \, \frac12 \left. \left( \rho \, \bar{u^2} \, \bar{u} \, r \, h \right) \right|_{r+\Delta r} 
\, - \, \left. \left( \gamma \, \bar{u} \, r \right) \right|_r 
\, + \, \left. \left( \gamma \, \bar{u} \, r \right) \right|_{r+\Delta r} \, + \cdots \ = \ 0 \ ,
\end{equation}

\noindent
where $\bar{u} \, = \, C_{1} \, u_{\mathrm{s}}$ and $\bar{u^{2}} \, = \, C_{2} \, u_{\mathrm{s}}^{2}$ denote averages. The two terms containing $\gamma$ are new and introduced without reference to represent the flow of surface energy. The two terms are incorrect, and we now proceed to derive the appropriate terms. 

We shall build our derivation on the standard energy equation (see, e.g., \citet[Eq. (16.2)]{Landau1987}). For time-independent Newtonian and incompressible flows the equation reduces to

\begin{equation}
\label{eq:energy-basic}
 - \oint \left[ v_j \left( \frac12 \rho \, v^2 + p \right) -  \, \mu v_i v_{ij} \right] n_j \, \mathrm{d}A \, - \, \frac12 \, \mu \int v^{2}_{ij} \, \mathrm{d} V \ = \ 0 \ .
\end{equation}

\noindent 
To include surface tension, we use the conclusion of the previous section given by Eq. (\ref{eq:normal}): the only effect due to surface tension is that the pressure at the surface contains an additional component, the Laplace pressure, $p_{\gamma}$, given by Eq. (\ref{pgamma}).

As usual in thin film flows \cite{Oron1997}, we approximate the pressure as the sum of the Laplace term and a hydrostatic pressure term:

\begin{equation}
p(r,z) \ = \ p_{0} \, + \, \gamma \, \kappa(r) \, + \, \rho \, g \left[ h(r)-z \right] \ ,
\end{equation}

\noindent
where $\kappa$ for a surface of revolution is given by

\begin{equation}
\label{kappa}
\kappa \ = \ - \frac{1}{r}\frac{\mathrm{d}}{\mathrm{d}r}\left( \frac{r h'}{\sqrt{1+ (h')^2}}\right) \ .
\end{equation}

\noindent
In principle, the term $p_{\mu}$ should also be included, but typically it is not \cite{Watson1964}. In any case it is unrelated to surface tension, and if it is important, it should  be included in the analysis of the viscous flow without surface tension, i.e., in the $p$ appearing in Eq. (\ref{eq:energy-basic}). We shall later give an estimate of its magnitude for hydraulic jumps.

We shall focus on the kinetic energy and pressure surface integral term in Eq. (\ref{eq:energy-basic}), and, as \citet{Bhagat2018}, we consider an annular control volume (Fig. \ref{fig-geometry}). The surface integral over the inner cylindrical surface leads to a sum of three terms representing the kinetic energy, the Laplace pressure, and the hydrostatic pressure, respectively. Setting $p_0 = 0$ and omitting a factor of $2 \, \pi$ we find the expression

\begin{equation}
\label{ener2}
\chi(r) \ = \ \int_0^h  u \left( \frac12 \, \rho \, (u^2 + w^2)+ p\right) \, r \, \mathrm{d}z
\ \approx \ \frac12 \, C_3 \, \rho \, r \, u_{\mathrm{s}}^{3} \, h
\, + \, \gamma \, q \, \kappa 
\, + \, C \, \rho \, g \, r \, u_{\mathrm{s}} \, h^{2} \ ,
\end{equation}

\noindent
where $q$ is the conserved flow rate per radian
\begin{equation}
\label{eq:flow-rate}
  q \, = \, C_{1} \, r \, u_{\mathrm{s}} \, h \ .
\end{equation}
Here we have omitted the $w^2$-term as done (without comment) by \citet{Bhagat2018}. Since they are interested in small $h'$, this seems reasonable as $w \sim h' u$ from Eq. (\ref{eq:profilev}). The surface tension term $ \gamma \, q \, \kappa $ in Eq. (\ref{ener2}) is only non-zero if the surface is curved, and it should be contrasted with the term $\gamma \, \bar{u} \, r =\gamma q/h $ postulated by \citet{Bhagat2018}. From the outer cylindrical surface we obtain an integral similar to Eq. (\ref{ener2}), but with a minus sign, and we proceed to derive the differential energy equation. The Laplace pressure term will give us up to third order derivatives of $h$. In the spirit of \citet{Bhagat2018} and \citet{Mathur2007} we restrict our attention to the lowest (first) order derivatives and the lowest order (linear) terms in $h'$. We obtain $\kappa \, \approx \, - h'/r$ and

\begin{equation}
\label{eq:kappa-approximation}
\frac{\partial p_{\gamma}}{\partial r}=\gamma \kappa'(r) \ \approx \ \gamma \frac{h'(r)}{r^2} \ .
\end{equation}

\noindent
This approximation would be entirely wrong near the contact line outside the jump, where the $h'''$ term dominates, but inside the jump it might be reasonable \cite{Mathur2007}. It corresponds to neglecting the curvature in the vertical $rz$-plane, retaining only the curvature that captures the difference between a jump in a channel and the circular one discussed here. 

Using the approximation (\ref{eq:kappa-approximation}) and that $h' \, = \, - h \, (1/r \, + \, u_{\mathrm{s}}'/u_{\mathrm{s}})$ since the flow rate per radian $q$ is constant, we obtain the expression

\begin{equation}
\label{chi'}
  \chi' \ \approx \ 
  \left( C_{3} \, \rho \, u_{\mathrm{s}}^{2} \, h \, - \, 
  C_{1} \, \alpha^{2} \, \gamma \, - \, C \, \rho \, g \, h^{2} \right) r \, u_{\mathrm{s}}' \, - \, \left(
  C_{1} \, \alpha^{2} \, \gamma \, + \, C \, \rho \, g \, h^{2} \right) u_{\mathrm{s}} \ ,
\end{equation}

\noindent
where we have collected the terms involving the derivative of the radial velocity at the free surface. Finally, we can write the differential energy equation

\begin{equation}
\label{eqf}
 u_{\mathrm{s}}' \ \approx  \ \frac{\left(
  C_{1} \, \alpha^{2} \, \gamma \, + \, C \, \rho \, g \, h^{2} \right) u_{\mathrm{s}} \, + \, \xi}{
 \left[ 1 \, - \, (C_{1}/C_{3}) \, \alpha^{2} \, \mathrm{We}^{-1}
 \, - \, (C/C_{3}) \, \mathrm{Fr}^{-2}
 \right] C_{3} \, \rho \, r \, u_{\mathrm{s}}^{2} \, h} \ ,
\end{equation}

\noindent
where we have made use of the dimensionless numbers defined in Eq. (\ref{eq:dimensionless}), and where $\xi$ represents the viscous terms from Eq. (\ref{eq:energy-basic}). Except for numerical factors of order unity, we observe that the expression becomes singular when the condition (\ref{R2}) is satisfied. Our result is to be contrasted with the differential energy equation derived by \citet[Eq. (5.6)]{Bhagat2018}, which instead, again except for numerical factors of order unity, will become singular when the condition (\ref{R1}) is satisfied.

\section{The influence of viscosity}

The viscous pressure term $p_{\mu} $ that is usually neglected in thin film flows can be determined from Eq. (\ref{pmu}). Using ${\bf n} \, = \, [1+(h')^2]^{-1/2} \, (-h',1)$, we find
\begin{align}
\label{pmuA}
p_{\mu} \ = \  \frac{2 \, \mu}{\left[1+ (h')^2 \right]} \left[ (h')^2 \, \frac{\partial u}{\partial r} \, - \, h' \left(\frac{\partial u}{\partial z} \, + \, \frac{\partial w}{\partial r}\right) \, + \, \frac{\partial w}{\partial z} \right] \ ,
\end{align} 
where all derivatives are evaluated on the free surface $z \, = \, h(r)$. The corresponding no-stress condition ({\ref{eq:tangential}) is:
\begin{align}
\label{no-stress}
 [1 - (h')^2]\left(\frac{\partial u}{\partial z}+\frac{\partial w}{\partial r}\right) \ = \ 2 \, h' \left(\frac{\partial u}{\partial r} -\frac{\partial w}{\partial z}\right) \ .
\end{align}
In the usual treatment of hydraulic jumps \cite{Watson1964}, the viscous pressure $p_{\mu}$ is neglected and the stress condition (\ref{no-stress}) is approximated for small $h'$ simply as $\partial u/ \partial z=0$. Accepting the latter approximation we can estimate the viscous pressure as
\begin{equation}
\label{pmu1}
p_{\mu} \ \approx \ 2 \, \mu \left.\frac{\partial w}{\partial z}\right|_{z=h} \ \approx \ 2 \, \mu \, u_s \, \frac{h'}{h} \ = \ \frac{2 \, \mu \, q \, h'}{C_1 \, r \, h^2} \ .
\end{equation}
For the last transformation we have used the vertical velocity profile  (\ref{eq:profilev}). Thus (using $p_{\gamma} \approx - \gamma h'/r$) we find
\begin{equation}
\label{pratio}
\frac{p_{\mu}}{p_{\gamma}} \ \approx \ - \frac{2 \, \mu \, q}{C_1 \gamma \, h^2}.
\end{equation}
Taking, e.g., the surface profile for a circular hydraulic jump in ethylene glycol mixed with water \cite{Bohr1996}, we have
$\mu \approx 10^{-2} \, \mathrm{Pa} \, \mathrm{s}$, $\gamma \approx 50 \, \mathrm{mN} \, \mathrm{m}^{-1}$, $h \, \approx \, 1 \, \mathrm{mm}$, $Q \, = \, 2 \, \pi \, q \, = \, 30 \, \mathrm{cm^3} \, \mathrm{s}^{-1}$ and $C_1 \approx 0.62$ (\citet{Watson1964}), and we find $p_{\mu} \, \approx \, -3.1 \, p_{\gamma}$. Thus, in this case, the magnitude of the viscous contribution is approximately three times as big as the capillary one.

\section{Concluding remarks}

It is interesting to note that the term $\gamma \, r \, \bar{u} $ postulated by \citet{Bhagat2018} in Eq. (\ref{E1}) corresponds to adding a pressure
\begin{equation}
\label{pB}
p_B \ = \ - \, \gamma \, \frac{1}{h} \ .
\end{equation}
In the Navier-Stokes equation, the driving force is provided by (minus) the gradient of the pressure, i.e., 
\begin{equation}
\label{gradpB}
\frac{\partial p_B}{\partial r} \ = \ \gamma \, \frac{h'}{h^2} \ ,
\end{equation}
which should be contrasted with the result for small $h'$ in Eq. (\ref{eq:kappa-approximation}). The relation $\partial p_{\gamma}/\partial r \, \approx \, \alpha^{2} \, \partial p_{B}/\partial r$ explains the origin of the factor $\alpha^2$ in Eq. (\ref{R2}) and the huge overestimate of the surface tension effect made by \citet{Bhagat2018}. 

We would like to make clear that we are not postulating that the radius of the circular hydraulic jump should satisfy the condition (\ref{R2}). The radius of the jump in an expanding, time-dependent flow that has not reached the outer rim can not necessarily be inferred from studying the stationary system, nor by studying only the {\em inner} flow. The jump signifies a transition between an inner and an outer flow, and, as emphasized already by B{\'e}langer and Rayleigh, the location of the jump depends on both states. What we can hope to achieve by investigating the existence of a non-singular inner flow is thus only an {\em upper bound} on the radius of the jump. As discussed in the introduction, the energy equations used by \citet{Bhagat2018} and in the present paper neglect any time-dependence. During the initial formation of a jump - or, at least initially, a {\em rim} - this is not justified, and we believe that the introduction and evaluation of such terms is a worthwhile direction for further studies. Finally, the viscous pressure is a hitherto neglected effect that, as the estimate in Eq. (\ref{pratio}) shows, can be of similar magnitude as that of surface tension.\\

The research leading to these results has received funding from the People Program\-me (Marie Curie Actions) of the European Union's Seventh Framework Programme (FP7/2007-2013) under REA grant agreement no. 609405 (COFUND Postdoc DTU). We thank Mederic Argentina for comments on the initial manuscript.


\end{document}